\documentclass[preprint2]{aastex}
\usepackage{amsmath}
\usepackage[usenames,dvipsnames]{xcolor}
\usepackage{multirow}

\shorttitle{BH mass estimation for 1H~0323+342}
\shortauthors{Pan et al.}

\begin{document}

\title{Independent Estimation of Black Hole Mass for the $\gamma$-ray Detected Archetypal Narrow-Line Seyfert 1 Galaxy 1H~0323+342 from X-ray Variability}

\author{Hai-Wu Pan\altaffilmark{1,2}, Weimin Yuan\altaffilmark{1,2}, Su Yao\altaffilmark{3,4}, S. Komossa\altaffilmark{5} \& Chichuan Jin\altaffilmark{1}}

\altaffiltext{1}{Key Laboratory of Space Astronomy and Technology, National Astronomical Observatories, Chinese Academy of Sciences, 20A Datun Road, Chaoyang District, Beijing, China; panhaiwu@nao.cas.cn, wmy@nao.cas.cn}
\altaffiltext{2}{University of Chinese Academy of Sciences, School of Astronomy and Space Science, Beijing 100049, China}
\altaffiltext{3}{Kavli Institute for Astronomy and Astrophysics, Peking University, Beijing 100871, China}
\altaffiltext{4}{National Astronomical Observatories, Chinese Academy of Sciences, Beijing, China}
\altaffiltext{5}{Max-Planck Institut f{\"u}r Radioastronomie, Auf dem H{\"u}gel 69, 53121 Bonn, Germany}

\begin{abstract}
$\gamma$-ray detected narrow-line Seyfert 1 galaxies (NLS1s) are a newly discovered class of radio-loud active galactic nuclei (AGNs) that launch powerful jets which are generally found only in blazars and radio galaxies. However, their black hole (BH) masses as estimated from their broad emission lines are an order of magnitude or more lower than those in blazars. This poses new challenges in explaining the triggering of radio loudness in AGNs. It is still under debate whether their BH masses are underestimated by the commonly used virial method. Here we present an estimate of the BH mass for the $\gamma$-ray detected NLS1 1H~0323+342, an archetype of this class, from its X-ray variability, which is independent of inclination. Our results independently confirm that this $\gamma$-ray detected NLS1 harbors a $(2.8-7.9)\times10^6\,M_{\odot}$ BH similar to those in normal NLS1s rather than those in blazars.
\end{abstract}

\keywords{galaxies: active --- galaxies: nuclei --- X-rays: galaxies}

\section{INTRODUCTION}

Narrow-line Seyfert 1 galaxies (NLS1s), which are defined as active galactic nuclei (AGNs) with broad Balmer lines narrower than 2000\,km\,s$^{-1}$ \citep{O1985}, appear mostly to be radio-quiet with radio loudness\footnote{The radio loudness $R$ is defined as the ratio of the radio 5\,GHz flux to the optical B-band flux.} $R<10$. Only a small subset have been noticed to be radio-loud sources (e.g., \citealp{Z2006,K2006}). \citet{Y2008} found there exists a population of genuine radio-loud NLS1s showing observational properties characteristic of blazars with relativistic jets. The finding was confirmed and highlighted by the detection of $\gamma$-ray emission from a small number of NLS1s with the Fermi satellite (e.g., \citealp{A2009a,A2009b,D2012,Y2015a}). Radio-loud NLS1s exhibit some distinct properties compared to those classical radio-loud AGNs, e.g., blazars. Nearby NLS1s generally reside in pseudo-bulges \citep{M2012}, while the common hosts of classical radio-loud AGNs are elliptical galaxies (e.g., \citealp{L2011}). The key difference is the mass of the central black hole (BH).

NLS1s are thought to have lower BH masses than classical blazars and Seyfert galaxies, and hence higher Eddington ratios (see \citealp{K2008,Y2008} for a review). However, this assertion is still under active debate at the forefront of research today. There have been studies suggesting that the BH masses of NLS1s may possibly be underestimated by the commonly used virial method involving broad optical emission lines because of the inclination effect, if their broad line regions (BLRs) are planar and seen face on \citep{C2004,J2006}.
Early studies have found a positive correlation between the radio loudness and BH mass in AGNs \citep{L2000}. This indicates that a threshold BH mass is required for AGNs that launch powerful relativistic jets. Thus, radio-loud NLS1s have also been suggested in some studies to have supermassive BHs similar to those in blazars (e.g., \citealp{C2013,L2014,B2016,D2017,D2018}).
Moreover, observationally, NLS1s are found to be located at one extreme end in the so-called eigenvector 1 parameter space \citep{B2002,S2008}, as revealed by a set of correlations presumably driven primarily by the Eddington ratio (e.g., \citealp{B2002,G2010,X2012}). These include the strong permitted Fe emission lines \citep{B1992}, steep soft X-ray spectra \citep{W1996} and rapid X-ray variability \citep{L1999}, etc. In contrast, most blazars lie at the other end of the eigenvector 1 parameter space \citep{B2002}. Thus, NLS1s were once thought to be radio-quiet, which turns out to be a consequence of their low radio-loud fraction \citep{Z2006,K2006}. Thus the detection of $\gamma$-ray emitting NLS1s poses a new challenge in explaining the radio loudness in AGNs.
There are other lines of evidence hinting that radio-loud NLS1s have lower BH masses than blazars, and hence higher Eddington ratios \citep{K2006,Y2008,Y2015b}.
Therefore, an inclination-independent estimator is essential for resolving the controversy over the BH masses of NLS1s that launch relativistic jets.

Only a handful of NLS1s have been detected in $\gamma$-rays so far \citep{A2009a,A2009b,F2011,D2012,D2015,Y2015a,P2018,Y2018}. 1H~0323+342 is the nearest of them (redshift $z=0.0629$), allowing detailed observational studies, e.g., of its host morphology \citep{Z2007,L2014}, the variability of multiwavelength emission (e.g., \citealp{Pa2014,Y2015b}), and jet structure on the scale of parsecs \citep{Fuh2016}.
The method of single-epoch spectra using several broad emission lines has been employed, resulting in a relatively low mass of about $\sim2\times10^7\,M_{\odot}$ \citep{Z2007,L2017}, and an Eddington ratio of about $0.5-1.0$ \citep{L2017}. Recently, \citet{W2016} used the reverberation mapping method and obtained a similar result. However, a BH mass of the order of magnitude of $10^8\,M_{\odot}$ has also been suggested, based on relation between bulge luminosity and BH mass \citep{L2014}.
This discrepancy can be explained in two ways: either 1H~0323+342 (and perhaps other radio-loud NLS1s too) resides in a luminous bulge (often a pseudo-bulge) that harbors a much less massive BH than what is normally predicted from the bulge luminosity, or the BH mass is underestimated by the virial method using the broad lines as a result of the inclination effect.

In this paper, we estimate the BH mass of 1H~0323+342 from the variability properties of its X-ray emission. Rapid X-ray variability is one of the basic observational characteristics of AGNs \citep{M1985}. Since the X-ray emission is thought to originate from the innermost region of an accretion flow around the BH, its variability provides a powerful tool to study the dynamics of matter very close to the BH, which is likely encoded with some of the BH parameters. In fact, some of the characteristics of X-ray variability have been used to estimate the BH mass of AGNs, including the break frequency of the power spectral density (PSD) \citep{M2006,G2012}, the normalized excess variance \citep{Z2010,P2012}, and the quasi-periodic oscillation frequency \citep{P2014,Z2015,P2016}. Unlike the commonly used virial method involving the line-of-sight velocities of the orbiting gas, which is susceptible to the orientation effect of AGNs \citep{C2004}, X-ray variability can be considered as essentially independent of inclination. In this work, we perform X-ray timing analysis using data from a high quality X-ray observation made with the {\it XMM-Newton}, based on which the BH mass of 1H~0323+342 is derived from the break frequency of the PSD and the excess variance.

\section{OBSERVATION AND DATA REDUCTION}

1H~0323+342 was observed with {\it XMM-Newton} with an exposure of $\sim80$\,ks on 2015 August 24 in the large window imaging mode (ObsID: 0764670101).
To improve the signal-to-ratio, the data taken from the EPIC PN and MOS cameras are utilized. We follow the standard procedure to reduce the data and extract science products from the observation data files (ODFs) using the {\it XMM-Newton} Science Analysis System (version 15.0.0). Only good events (single and double pixel events, i.e., PATTERN $\leq $ 4 for PN or $\leq $ 12 for MOS) are used. Source events are extracted from a circular region of 40 arcsec radius, and background events from a source-free circle with the same radius on the same CCD chip. No pile-up effect is found for each of the detectors, by applying the EPATPLOT task. X-ray light curves are constructed for all the three EPIC cameras in the 0.2-10\,keV energy band with a binsize of 10\,s. Final source light curves are obtained by subtracting corresponding background light curves extracted from the background regions, and corrected for instrumental factors using EPICLCCORR.

\begin{figure}
\plotone{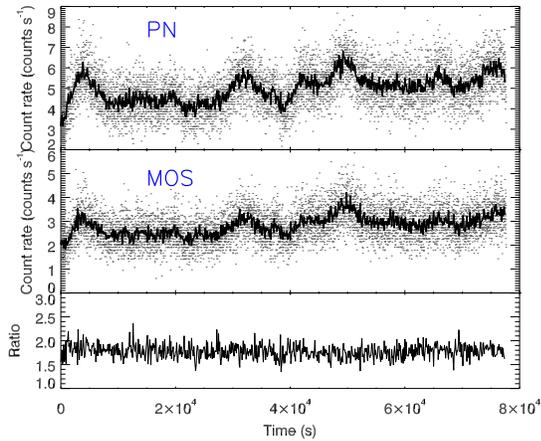}
\caption{The PN (upper panel) and combined MOS (MOS1+MOS2; middle panel) light curves of 1H~0323+342 in the energy band of 0.2-10\,keV. The black lines represent the light curves with a binsize of 100\,s, while those with a binsize of 10\,s are plotted in gray. The ratio between the PN and combined MOS (MOS1+MOS2) light curves is plotted in the lower panel. \label{fig1}}
\end{figure}

\begin{figure}
\plotone{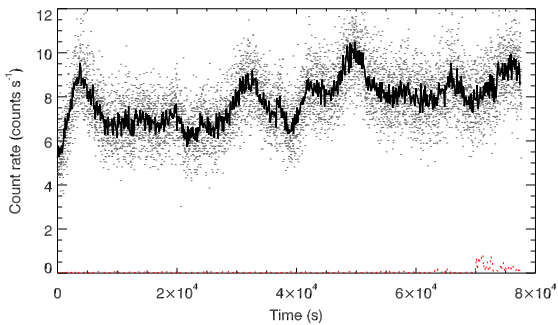}
\caption{X-ray light curve of 1H~0323+342 obtained with {\it XMM-Newton}, which is a combination of background-subtracted light curves extracted from the PN, MOS1 and MOS2 detectors in 0.2-10\,keV. The black line represents the light curve with a binsize of 100\,s, while the one with a binsize of 10\,s is plotted in gray. The background (red dashed line) is significantly fainter than the source. \label{fig2}}
\end{figure}

Count rate light curves of a source measured simultaneously by independent detectors can be co-added in order to increase the signal-to-noise ratio, provided that the detectors have the same responses (effective areas) or the source's spectral shape does not vary with time during the observation. Since the two MOS detectors are essentially identical and have very similar responses, the two MOS light curves are co-added into a combined MOS light curve. Figure 1 shows the PN and the combined MOS light curves. Although the PN and MOS detectors have different responses, the two light curves follow each other closely, indicating weak or no changes in the spectral shape. This can be demonstrated by the ratio of the two light curves (bottom panel in Figure\,\ref{fig1}), which appears to be constant. We thus also co-added the PN and MOS light curves (CR$_{\rm total}$=CR$_{\rm PN}$+CR$_{\rm MOS1}$+CR$_{\rm MOS2}$) to form a combined single PN+MOS light curve of the source, assuming no significant spectral variations during the observation. The combined light curve for all the detectors is shown in Figure\,\ref{fig2}. In the following analysis, both the PN and MOS light curves are used individually, as well as the combined single source light curve with the best signal-to-noise ratio for comparison.
As shown in Figure\,\ref{fig1} and Figure\,\ref{fig2}, the X-ray emission of 1H~0323+342 exhibits significant variability on various timescales, typical of NLS1 galaxies.

During the whole observation the particle background is at a normal level except for the last $\sim$15\,ks when an enhanced background flare occurred.
We have carefully inspected the data set for a possible influence of the background on the measured light curves, and especially during the intervals of the flaring background. We find that the background is significantly fainter (mean count rate=0.05\,counts\,s$^{-1}$; Figure\,\ref{fig1}) than the source, which is rather bright (mean count rate=7.80\,counts\,s$^{-1}$). Even during the period of enhanced background toward the end of the observation, the background is more than five times fainter than the source for each time bin.
We therefore consider the effect of background fluctuations on the measured light curves to be negligible, and thus make use of the full, uninterrupted source light curve for timing analysis.

\section{TIMING ANALYSIS AND ESTIMATION OF BH MASS}

\subsection{Power Spectral Density}

Early studies showed that the PSD of AGNs could be described by a simple power law \citep{L1993}, whereas later investigations with high quality PSDs revealed a flattening toward the low frequency end (e.g., \citealp{P1995}). It has been established that the AGN PSD can be best described by a bending (or broken) power law, with two slopes with universal values of $\sim -2$ in the high frequency part and $\sim -1$ in the low frequency part\citep{M2003,V2003a}. \citet{M2006} first suggested a tight correlation between the bend/break frequency $\nu_{\rm b}$ and the BH mass as well as the bolometric luminosity over a wide range of BH masses from black hole X-ray binary (BHXB) to AGN. Using a larger sample of AGNs, \citet{G2012} found a similar result, but with a weaker dependence on the bolometric luminosity. This indicates that the bend/break frequency depends mostly on the BH mass.

\begin{figure}
\plotone{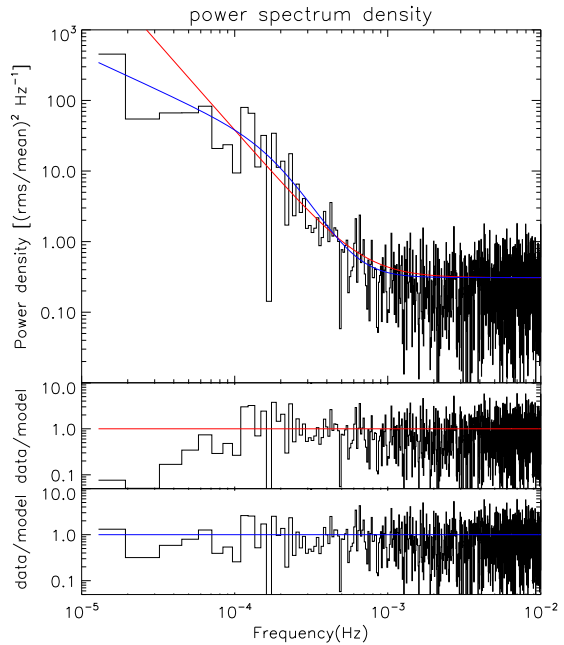}
\caption{Power spectral densities of the combined PN+MOS X-ray light curve of 1H~0323+342. The red solid line represents the best-fit power-law, while the blue solid line donates the best-fit bending power-law. The data/model residuals for power-law and bending power-law models are shown in the middle and bottom panel, respectively. \label{fig3}}
\end{figure}

In the analysis below, the combined PN+MOS light curve is mainly used, given its highest signal-to-noise ratio, while the PN and the combined MOS light curves are also analyzed for the purpose of comparison of results.
The PSDs of the light curves of 1H~0323+342 are computed as the modulus-squared of the discrete Fourier transform, and the (rms/mean)$^2$ normalization is chosen \citep{V2003b}. The PSD derived from the combined PN+MOS light curve is shown in Figure\,\ref{fig3}. It shows a typical red-noise spectrum in the frequency range from $10^{-5}$ to $10^{-3}$\,Hz, while above $10^{-3}$\,Hz it is dominated by Poisson noise.
As can be seen, a flattening toward the low frequency end is apparent in the red noise of the PSD.

We first fit the PSD with a model of a simple power-law (red noise) plus a constant (Poisson noise): $P(\nu)=N\nu^{-\alpha}+C$ (\citealp{G2012}, see also the $H_{0}$ model in \citealp{V2010}), where the normalization $N$, power-law index $\alpha$ and additive constant $C$ are all set to be free parameters. The maximum likelihood method is used to find the best-fitting model parameters by minimizing the fit statistic $S$ which is (twice) the minus log-likelihood (\citealp{G2012}; see also \citealp{V2010}), yielding a slope of $2.47\pm0.18$ (the red line in Figure \ref{fig3}). The uncertainty of each parameter can be estimated by finding the range of parameter values for which $\Delta S\leq 1.0$, corresponding to a significance level of 68.3\% \citep{C1979,G2012}.
However, a single power-law does not describe the PSD well in the low frequency regime. Judging from the data/model residuals (see the middle panel of Figure \ref{fig3}), the slope of the PSD appears to flatten at around a frequency of $10^{-4}$\,Hz.
Thus we adopt a bending power-law ($P(\nu)=N\nu^{-1}(1+\left\{\frac{\nu}{\nu_b}\right\}^{\alpha-1})^{-1}+C$) (\citealp{G2012}, see also the $H_{1}$ model in \citealp{V2010}) instead of a simple power-law to fit the PSD. The low-frequency slope is fixed to be the canonical value -1 for AGNs found from long-term X-ray monitoring studies (e.g., \citealp{M2003,M2006}), given the relatively poor data quality at the low-frequency end \citep{G2012}. The high-frequency slope is fitted to be $3.66\pm0.54$, with a bend frequency of $1.9\times10^{-4}$\,Hz.
Hence, the likelihood ratio test (LRT) suggested in \citet{V2010} (see also \citealp{G2012}) is used to select between the simple power-law and bending power-law models, giving a posterior predictive $p$-value $p=9\times10^{-5}$ from the Markov Chain Monte Carlo (MCMC) simulations \footnote{As outlined in \citet{V2010}, 100,000 sets of parameters have been simulated from the Metropolis-Hastings MCMC algorithm, based on the best-fitting (simple power-law) model parameters (and their errors). Each set of simulated parameters can be used to generate a periodogram. Then the two models (simple power-law and bending power-law) are fitted to the same simulated periodogram to obtain the LRT statistic. Here the LRT statistic is defined as the difference between twice the minimum log-likelihood of the two models. Thus we can get the posterior predictive $p$-value by comparing the LRT statistic for the observed data with the distribution of this statistic obtained from the simulated data. See \citet{V2010} and the appendices therein for details.}. This indicates that the PSD for 1H~0323+342 is well described by a bending power-law, as commonly found in Seyfert AGNs.

\begin{figure}
\plotone{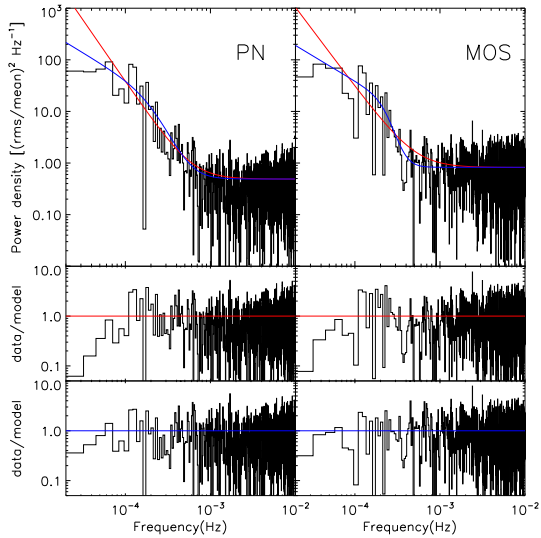}
\caption{Power spectral densities of the PN and combined MOS light curves of 1H~0323+342. The red solid lines represent the best-fit power-law, while the blue solid lines donate the best-fit bending power-law. The data/model residuals for power-law and bending power-law models are shown in the middle and bottom panels, respectively. \label{fig4}}
\end{figure}

We also analyze the PSDs of the PN and combined MOS light curves. Both PSDs can be fitted better by a bending power-law better, with a $p$-value of $p=3.2\times10^{-4}$ (PN) or $p=2.0\times10^{-4}$ (MOS) based on the LRT method, respectively, as shown in the Figure\,\ref{fig4}. The fitted bend frequencies are $2.1\times10^{-4}$\,Hz (PN) and $2.3\times10^{-4}$\,Hz (MOS), respectively, which are consistent with each other as well as that obtained from the combined PN+MOS light curve. Thus, the bend frequency of $1.9\times10^{-4}$\,Hz obtained from the combined light curve is used to estimate the BH mass of 1H~0323+342 hereinafter.
Besides, we also fit the PSD of 1H~0323+342 with a broken power-law model, given that the break frequency may be slightly different from the bend frequency, as discussed in \citet{V2010}. The fitted break frequency is $2.0\times10^{-4}$\,Hz, which is consistent with the bend frequency derived above.

The relation between the bend/break frequency and the BH mass as well as the bolometric luminosity was first suggested by \citet{M2006}. Later \citet{G2012} have found that the relation depends weakly on the bolometric luminosity, thus a relation only between the bend/break frequency and the BH mass is also suggested. All these relations are considered for the BH estimations of 1H~0323+342, and the results are listed in Table\,\ref{tbl1}.
First, we derive the BH mass for 1H~0323+342 from the bend frequency of $1.9\times10^{-4}$\,Hz using the relation with BH mass only:
\begin{equation}
\log(T_{\rm b})=(1.09\pm0.21)\log(M_{\rm BH})+(-1.70\pm0.29),
\end{equation}
where $T_{\rm b}$ is the bend timescale ($T_{\rm b}=1/\nu_{\rm b}$) \citep{G2012}\footnote{$T_{\rm b}$ is in units of days, $M_{\rm BH}$ in units of $10^6\,M_{\odot}$, and $L_{\rm bol}$ in units of $10^{44}$\,ergs\,s$^{-1}$ for the equations from \citet{M2006} and \citet{G2012}.}.
The BH mass is estimated to be $2.8\times10^6\,M_{\odot}$. The uncertainty range is calculated from Monte Carlo simulations, in which the uncertainties of the parameters in the correlations \citep{M2006,G2012} as well as $\log(T)$ and $\log(M_{\rm BH})$ are assumed to follow a Gaussian distribution.\footnote{The same method is also used for the following estimations for the uncertainty of BH mass. For some parameters as well as measurements, the errors are not symmetric. In this case, the relatively larger value of error will be used. Thus the uncertainties will be a little overestimated.} For $\log(T)$ with a value of $-1.22^{+0.15}_{-0.11}$, the uncertainty is computed via the propagation of error from the bend frequency.

Next we estimate the BH mass for 1H~0323+342 using the relation that depends on both the BH mass and bolometric luminosity.
Using the bolometric correction factor \citep{E1994}, \citet{Z2007} obtained a bolometric luminosity of $1.2\times10^{45}$\,ergs\,s$^{-1}$ for 1H~0323+342 from its luminosity at 5100\,\AA. \citet{W2016} monitored 1H~0323+342 for more than two months using the Lijiang 2.4\,m Telescope, and obtained a mean $L_{\rm 5100}$ of $7.52\times10^{43}$\,ergs\,s$^{-1}$. The bolometric luminosity can be estimated as $9L_{\rm 5100}$ by making use of the bolometric correction factor suggested in \citet{K2000}. The value $6.77\times10^{44}$\,ergs\,s$^{-1}$ is nearly consistent with the result in \citet{Z2007}, indicating that the optical variability of the source is weak on a timescale of about 10 years.
We make use of the relationship suggested in \citet{M2006}:
\begin{equation}
\begin{split}
\log(T_{\rm b})=(2.17^{+0.32}_{-0.25})\log(M_{\rm BH})&+(-0.90^{+0.3}_{-0.2})\log(L_{\rm bol})\\
&+(-2.42^{+0.22}_{-0.25})
\end{split}
\end{equation}
as well as the relationship in \citet{G2012}:
\begin{equation}
\begin{split}
\log(T_{\rm b})=(1.34\pm0.36)\log(M_{\rm BH})&+(-0.24\pm0.28)\log(L_{\rm bol})\\
&+(-1.88\pm0.36).
\end{split}
\end{equation}
The BH mass of 1H~0323+342 is estimated to be $7.9\times10^6\,M_{\odot}$ (Equation\,2) or $4.4\times10^6\,M_{\odot}$ (Equation\,3), using the bolometric luminosity of $6.77\times10^{44}$\,ergs\,s$^{-1}$ \citep{W2016}. The BH mass will be $1.0\times10^7\,M_{\odot}$ and $4.8\times10^6\,M_{\odot}$ if the bolometric luminosity of $1.2\times10^{45}$\,ergs\,s$^{-1}$ \citep{Z2007} is used, which is consistent with the former results.

\begin{deluxetable}{ccc}
\tabletypesize{\scriptsize}
\tablecaption{Estimated BH mass for 1H~0323+342 \label{tbl1}}
\tablewidth{0pt}
\tablehead{\colhead{Measurements}  & \colhead{$\log(M_{\rm BH}/M_{\odot})$} & \colhead{Reference}}
\startdata
\multicolumn{3}{c}{From the PSD bend frequency} \\
\hline
\hline
$f_{bend}=1.9^{+0.66}_{-0.49}\times10^{-4}$\,Hz    &  $6.4\pm0.3$         &   1  \\
\hline
$f_{bend}=1.9^{+0.66}_{-0.49}\times10^{-4}$\,Hz    &  $6.6\pm0.5$         &   1  \\
$L_{\rm bol}=6.77\times10^{44}$\,erg\,s$^{-1}$                          &                      &      \\
\hline
$f_{bend}=1.9^{+0.66}_{-0.49}\times10^{-4}$\,Hz    &  $6.9\pm0.2$         &   2   \\
$L_{\rm bol}=6.77\times10^{44}$\,erg\,s$^{-1}$                          &                      &       \\
\hline
\multicolumn{3}{c}{From the normalized excess variance of variability}  \\
\hline
\hline
${\sigma}_{\rm rms,40ks}^{2}=(1.45^{+0.64}_{-0.65})\times10^{-2}$       &    $6.9\pm0.2$       &   3 \\
\enddata
\tablecomments{Column 1: the BH mass estimators with the uncertainties at the 1\,$\sigma$ level; Column 2: BH mass derived from different methods for 1H~0323+342, with the uncertainties at the 1\,$\sigma$ level; Column 3: corresponding references, 1-\citet{G2012}, 2-\citet{M2006}, 3-\citet{P2012}.}
\end{deluxetable}

Finally, we calculate the PSDs of 1H~0323+342 separately for the soft-band light curve (0.2-1.0\,keV) and the hard-band light curve (2.0-10.0\,keV). We find that both PSDs show similar results to those presented above for the full band, albeit with lower significance of the existence of the bend/break frequency.

\subsection{Excess Variance}

Besides the PSD, the normalized excess variance of X-ray variations ($\sigma_{\rm rms}^2$) is also a useful and convenient tool to describe the variability amplitude. Following \citet{N1997}, the normalized excess variance can be calculated using the definition
\begin{equation}
{\sigma}_{\rm rms}^{2}=\frac{1}{N{\mu }^{2}}\sum_{i=1}^{N}\left[\left( {X}_{i}-\mu \right)^{2} - \sigma_i^2\right],
\end{equation}
where $N$ is the number of good time bins of an X-ray light curve, $\mu$ the unweighted arithmetic mean of the count rates, $X_i$ and $\sigma_i$ the count rates and their uncertainties, respectively, in each bin.

It has been shown that the tight inverse correlation between the X-ray excess variance and the BH mass of AGNs can be used to estimate the BH mass of AGNs \citep{L2001,P2004,Z2010}. \citet{P2012} reaffirmed this relationship by making use of a large sample of 161 AGNs observed with {\it XMM-Newton}, and found a small scatter of 0.4\,dex from the reverberation mapping sample. In fact, the $M_{\rm BH}-\sigma_{\rm rms}^2$ relation is actually a manifestation of the inverse scaling of the break frequency of the AGN PSD with the BH mass, since the excess variance is the integral of the PSD over the frequency domain \citep{P2004,P2015}.

We calculate the excess variances on a timescale of 40\,ks, using the light curve of 1H~0323+342 with a binsize of 250\,s and energy of 2-10\,keV from the PN camera only, which is the same as in \citet{P2012}. The resulting excess variance is 0.0145. A problem of this method is the large uncertainty, which comes from two sources: one is the measurement uncertainty and the other is the stochastic nature of the variability process, which is non-negligible \citep{V2003b,A2013}. The measurement uncertainty is calculated from the equation given in \citet{V2003b}, while the stochastic scatter is computed from the simulation method (see \citealp{V2003b,P2015} for more details), using the best-fit bending power-law PSD model (the Poisson noise is set to be zero) described in Section\,3.1. We simulate 1000 light curves using the method of \citet{T1995} with a binsize of 250\,s and duration of 40\,ks. The distribution of the calculated excess variances of all the light curves is obtained, from which the range of stochastic uncertainty at the significance level of 68\% is found: $\Delta\log(\sigma_{\rm rms}^2)=-0.25$ and $+0.16$. The total uncertainties are then obtained by combining in quadrature the stochastic and measurement uncertainties, which are at the $68\%$ confidence level.
The BH mass for 1H~0323+342 is derived using the relation suggested in \citet{P2012}:
\begin{equation}
\log(\sigma_{\rm rms}^2)=(-2.00\pm0.13)+(-1.32\pm0.14)\log(M_{\rm BH}/10^7\,M_{\odot}),
\end{equation}
yielding a BH mass of $7.5\times10^6\,M_{\odot}$.

In fact, it has been suggested that the $M_{\rm BH}-\sigma_{\rm rms}^2$ relation flattens at $\sim10^6\ M_{\odot}$, below which the excess variances tend to remain constant around 0.01 \citep{L2015,P2015}.
For 1H~0323+342, the value of excess variance is around 0.01, indicating that the BH mass is likely of the order of $\sim10^6\,M_{\odot}$ or even lower.

\section{SUMMARY AND DISCUSSION}

Thanks to the long, uninterrupted, and high quality X-ray observation with {\it XMM-Newton}, we are able to estimate the BH mass for the archetypal NLS1-blazar hybrid AGN 1H~0323+342 independently from the X-ray variability, which is free from the possible orientation effect of the BLR. The PSD of the light curve can be better fitted with a bending(broken) power-law  model than a simple power-law, with a bend frequency of $1.9\times10^{-4}$\,Hz. The previously established relationships between the PSD bend/break frequency and the BH mass (and possibly bolometric luminosity), as suggested by \citet{M2006} and \citet{G2012}, are employed.
The BH mass for 1H~0323+342 is constrained in the range of $(2.8-7.9)\times10^6\,M_{\odot}$, hence the Eddington ratio of $0.68-1.92$.
As an alternative approach, we also calculate the normalized excess variances of the light curve, giving a BH mass consistent with the results from the PSD break.
All these estimates are in reasonable agreement with one another, given the relatively large uncertainties that are calculated from Monte Carlo simulations (see Table\,\ref{tbl1}).

It should be noted that the BH mass---variability relationships used in this paper are based on studies of radio-quiet AGNs.
Although 1H~0323+342 is radio-loud, there is strong evidence (based on the X-ray spectral shape and multi-wavelength spectral energy distribution) that its X-ray emission below 10\,keV is predominantly from the disk/corona rather than from the relativistic jets, as in radio-quiet AGNs \citep{Y2015b}.
A similar conclusion was also reached by \citet{K2018}, who analyzed the same {\it XMM-Newton} observation used in this paper.
In fact, \citet{Y2015b} also calculated the X-ray normalized excess variance using a {\it Suzaku} observation with somewhat lower data quality and estimated a BH mass $\sim8.6\times10^6\,M_{\odot}$. The same practice using {\it Nustar} data was applied by \citet{L2017} and a similar result $\sim1.7\times10^7\,M_{\odot}$ was obtained.
These results are consistent with our estimates from the excess variance.
The BH mass derived independently from the X-ray variability for 1H~0323+342 is thus in agreement with that of $\sim2\times10^7\,M_{\odot}$ estimated from the broad optical/infrared emission lines \citep{Z2007,W2016,L2017}, which has a typical systematics of $\sim0.3$\,dex.

\citet{L2000} found that radio-loud AGNs tend to have BH masses larger than $3\times10^8\,M_{\odot}$. Furthermore, \citet{H2002} demonstrated that radio loudness is strongly inversely correlated with the Eddington ratio.
The small BH mass and high Eddington ratio found in 1H~0323+342 are at odds with the above findings. If the BH mass found for 1H~0323+342 is typical of radio-loud NLS1s with relativistic jets, this would imply a new parameter space in which relativistic jets can also be produced, i.e., AGNs with relatively small BH masses and high Eddington ratios (see \citealp{K2006,Y2008}). An interesting analogy was also suggested to BHXBs, which can also produce (transient) jets at the very high state when the Eddington ratios are the highest (e.g., \citealp{Y2008}).

Moreover, the consistency of the X-ray variability and virial estimation of the BH mass implies that the BLR in this NLS1 is not heavily flattened, because we have a near pole-on view of 1H~0323+342 as shown by \citet{Fuh2016}.
Given that 1H~0323+342 is just an individual case, more $\gamma$-ray emitting NLS1s with long, continuous and high quality X-ray observations are further needed to provide an independent estimations of BH mass.

\acknowledgments
This work is supported by the National Natural Science Foundation of China (Grant No.11473035). S. Y. thanks the support from the KIAA-CAS fellowship. This work is based on observations obtained with {\it XMM-Newton}, an ESA science mission with instruments and contributions directly funded by ESA Member States and NASA. This research has made use of the NASA/IPAC Extragalactic Database (NED).

\clearpage

\end{document}